\documentclass[aip,
jmp,amsmath,amssymb,
 reprint,%
]{revtex4-1}

\usepackage{graphicx}
\usepackage{dcolumn}

\newcommand{\al}{\alpha}
\newcommand{\bb}{\beta}

\newcommand{\ben}{\begin{eqnarray}}
\newcommand{\een}{\end{eqnarray}}
\newcommand{\be}{\begin{equation}}
\newcommand{\ee}{\end{equation}}
\newcommand{\n}{\label}
\newcommand{\no}{\noindent}

\newcommand{\ga}{\gamma}
\newcommand{\ro}{\rho}
\newcommand{\om}{\omega}

\newcommand{\bn}{\begin{equation}\label}

\begin{document}

\title{Holographic dark energy linearly interacting with dark matter}

\author{Luis P. Chimento}
\email[]{chimento@df.uba.ar}
\author{M\'onica I. Forte}
\email[]{monicaforte@fibertel.com.ar}
\author{Mart\'{i}n G. Richarte}
\email[]{martin@df.uba.ar}
\affiliation{Departamento de F\'isica, Facultad de ciencias Exactas y Naturales, Universidad de Buenos Aires, 1428 Buenos Aires, Argentina}

\date{\today}

\begin{abstract}We investigate a spatially flat Friedmann-Robertson-Walker (FRW) cosmological model with cold dark matter coupled to a modified holographic Ricci dark energy through a general interaction term linear in the energy densities of dark matter and dark energy, the total energy density and its derivative \cite{Chimento:2011dw}. Using the statistical method of $\chi^2$-function for the Hubble data, we obtain  $H_0=73.6$km/sMpc,  $\omega_s=-0.842$ for the asymptotic equation of state and   $ z_{acc}= 0.89 $. The estimated values of  $\Omega_{c0}$   which fulfill the current observational bounds  corresponds to a dark energy density varying in the range  $0.25R < \ro_x < 0.27R$.
\end{abstract}

\pacs{04.}
\maketitle

\bibliographystyle{plain}

\section{Introduction}

Many different observational sources such as the Supernovae Ia \cite{astro-ph/9805201}-\cite{astro-ph/9812133}, the large scale structure from  the Sloan Digital Sky survey \cite{arXiv:0707.3413} and the cosmic microwave background anisotropies \cite{arXiv:1001.4538}  have corroborated that our universe is currently undergoing an accelerated phase. The cause of this behavior has been attributed to a mysterious component called dark energy and several candidates have been proposed to fulfill this role. For example, a positive cosmological constant $\Lambda$, explains very well the accelerated behavior but it has a deep  mismatch with the  theoretical value predicted by the quantum field theory. Another issue of debate refers to the coincidence problem, namely: why the dark energy and dark matter energy densities happen to be of the same order precisely today. In order to overcome both problems, it has proposed a dynamical framework in which the dark energy varies with the cosmic time. This proposal  has led to a great variety of dark energy models such as quintessence \cite{astro-ph/9807002}, exotic quintessence \cite{arXiv:0706.4142},  N--quintom \cite{arXiv:0811.3643}  and the holographic dark energy (HDE) models \cite{hep-th/0403127} based  in an application of the holographic principle to the cosmology.    According to this principle, the entropy of a system does not scale with its volume but with its surface area and so in cosmological context will set an upper bound on the entropy of the universe \cite{hep-th/9806039}. It has been suggested \cite{hep-th/9803132} that in quantum field theory a short distance cut-off is related to a long distance cut-off (infra-red cut-off L) due to the limit set by the formation of a black hole. Further, if  the quantum zero-point energy density caused by a short distance cut-off is taken as the dark energy density in a region of size  L, it should not exceed black hole mass of the same size, so $\rho_{\Lambda}=3c^{2}M^{2}_{~P}L^{-2}$, where $c$ is a numerical factor. In the cosmological context, the size L is usually taken  as the large scale of the universe, thus Hubble horizon, particle horizon, event horizon or generalized IR cutoff. Among all the interesting holographic dark energy models proposed so far,  here  we focus our attention on a modified version of the well known Ricci scalar cutoff \cite{arXiv:0810.3663}.  Besides, there could be a hidden non-gravitational coupling between the dark matter and dark energy without violating current observational constraints and thus it is interesting to develop ways of testing an interaction in the dark sector. Interaction within the dark sector has been studied mainly as a mechanism to solve the coincidence problem.   We will consider an exchange of energy or interaction between dark matter and dark energy  which  is a linear combination of the dark energy density $\ro_x$,  total energy density $\ro$, dark matter energy density $\ro_c$, and the first derivate of the total  energy density  $\ro'$. \cite{arXiv:0911.5687}


\section{The interacting model}


In a FRW background, the Einstein equation for a model of cold dark matter of energy density $\rho_c$ and modified holographic Ricci dark energy having energy density $ \rho_x =\left(2\dot H + 3\alpha H^2\right)/\Delta$, reads 

\be
\n{1}
3H^2=\rho = \rho_c + \rho_x,
\ee 

\no where $\alpha$, $\beta$ are constants and $\Delta=\alpha -\beta$. 

In terms of the variable  $\eta = 3\ln(a/a_0)$, the  compatibility between the global conservation equation 
\be
\n{2}
\rho ' = d\rho/d\eta= -\rho_c-(1+\om_x)\rho_x, 
\ee
\no and the equation deduced from the expression of the modified holographic Ricci dark energy 
\be
\n{3}
\rho ' = -\alpha\rho_c-\beta\rho_x,
\ee
\no namely, $(\ro_c + \ga_x\ro_x)=(\al\ro_c + \bb\ro_x)$, gives a relation between the equation of state of the dark energy component $\om_x = \ga_x - 1$ and the ratio $r= \rho_c/\rho_x$
\ben
\n{4}
\om_x=(\alpha  - 1)r+\beta-1.
\een

Solving the system of equations (\ref{1}) and (\ref{3}) we get $\rho_c$ and $\rho_x$ in terms of $\ro$ and $\ro'$ as
\ben
\n{5}
\rho_c=-(\beta\rho+\rho')/\Delta,  \qquad   \rho_x=(\alpha\rho+\rho')/\Delta.
\een

The interaction between both dark components is introduced through the term $Q$ by splitting the Eq.(\ref{3}) into $\rho'_c+ \alpha \rho_c = - Q $ and  $ \rho'_x+ \beta \rho_x  =   Q $. Then, differentiating $\rho_c$ or $\rho_x$ in (\ref{5}) and using the expression of  $ Q $  we obtain a second order differential equation for the total energy density $\rho$   \cite{arXiv:0911.5687}

\be
\n{6}
\rho''+(\alpha+\beta)\rho'+\alpha\beta\rho = Q \Delta.
\ee

For a given interaction $Q$, solving Eq. (\ref{6}) gives us the total energy density $\ro$ and the energy densities $\ro_{c}$ and $\ro_{x}$ after using Eq. (\ref{5}).

The general linear interaction $ Q$ \cite{arXiv:0911.5687}, linear  in $\ro_{c}$, $\ro_{x}$, $\ro$, and $\ro'$, can be written as

\begin{eqnarray}
\label{7}
Q= c_1 \frac{(\gamma_s - \alpha)(\gamma_s-\beta)}{\Delta}\,\rho + c_2 (\gamma_s-\alpha)\rho_c \\
\nonumber  
- c_3 (\ga_s -\bb)\ro_x -c_4 \frac{(\ga_s - \al)(\ga_s-\bb)}{\ga_s\Delta}\,  \rho',
\end{eqnarray}

\no where $\ga_s$ is constant and  the coefficients $c_{i}$ fulfill the condition $c_{1}+c_{2}+c_{3}+c_{4}=1$ \cite{arXiv:0911.5687}.  

Now, using Eqs. (\ref{5}) we rewrite the interaction (\ref{7}) as a linear combination of $\ro$ and $\ro'$, 

\ben
\label{8}
Q=\frac{u\ro+\ga^{-1}_{s}[u-(\ga_s-\al)(\ga_s-\beta)]\ro'}{\Delta}, 
\een

\no where $u=c_1(\ga_s -\al)(\ga_s-\beta)-c_{2}\beta(\ga_s-\al)-c_{3}\al(\ga_s-\beta)$.  
Replacing the interaction (\ref{8}) into the source equation (\ref{6}), we obtain 

\be
\label{9}
\ro'' + (\ga_s+ \ga^{+})\ro'+ \ga_{s}\ga^{+}\ro=0.
\ee
\no where the  roots of the characteristic polynomial associated with the second order linear differential equation (\ref{9}) are  $\ga_{s}$ and $\ga^{+}=(\beta\al -u)/{\ga_s}$. 
In what follows, we adopt $\ga^{+}=1$ for mimicking the dust-like behavior of the universe at early times. In that case, the general solution of (\ref{9}) is   $\rho=b_1a^{-3\gamma_s}+b_2a^{-3}$  from which we obtain

\begin{subequations}
\n{10}
\begin{equation}
\n{10a}
\rho_c=\frac{(\gamma_s-\beta)b_1a^{-3\gamma_s}+(1-\beta)b_2a^{-3}}{\Delta},
\end{equation}
\begin{equation}
\n{10b}
\rho_x=\frac{(\alpha-\gamma_s)b_1a^{-3\gamma_s}+(\alpha-1)b_2a^{-3}}{\Delta}.
\end{equation}
\end{subequations}

Interestingly, Eqs. (\ref{10}) tell us that  the interaction (\ref{8}) seems to be a good candidate for alleviating the cosmic coincidence problem because the ratio $\Omega_{c}/\Omega_{x}$ becomes bounded for all times.

\subsection*{$Q= (1-\alpha)\rho_c $}

Let us consider the particular case in which the interaction $Q$ is proportional to the energy density of the dark matter $\rho_c$ in such a way that $\rho'_c+\rho_c =  \rho'_x+ (1+\om_x) \rho_x  = 0 $.  That is, each fluid separately, satisfies an equation of conservation. Here  the constants $u$ and $\gamma_s$ defined  above, correspond to $ u =\beta (\alpha-1)$  and  $\gamma_s = \beta $ with which the expressions (\ref{10}) for the energy densities of the dark matter and dark energy are written as functions of the redshift as
\begin{subequations}
\n{11}
\begin{equation}
\n{11a}
\rho_c=\frac{(1-\beta)b_2(1+z)^{3}}{\Delta},
\end{equation}
\begin{equation}
\n{11b}
\rho_x=\frac{(\alpha-\beta)b_1(1+z)^{3\beta}+(\alpha-1)b_2(1+z)^{3}}{\Delta}.
\end{equation}
\end{subequations}

The ratio between both components $r=\rho_c/\rho_x$ turns out to be

\begin{equation}
\n{12}
r=\frac{b_2(1-\beta)(1+z)^{3(1-\beta)}}{b_1\Delta + b_2(\alpha - 1)(1+z)^{3(1-\beta)}},
\end{equation}

\no and shows that in the early universe, both components behave as dust. In the final stages the ratio tends to zero and therefore it does not solve the problem of the coincidence. This example of interaction, between non-relativistic dark matter and the modified holographic Ricci dark energy, is important because allows to show that the holographic forms of the dark energy are always  interacting with the non-holographic component.  This behavior can be observed in the equation (\ref{11b}) and is due to the functional dependence of the holographic equation of state with the ratio $r$ of the energy densities, 

\ben
\n{13}
\om_x=\frac{(\beta-1) b_1 \Delta}{b_1 \Delta + b_2 (\alpha  - 1)(1+z)^{3(1-\beta)}} .
\een

each time $\alpha$ is different from 1.

\section{Observational constraints}

The transition redshift $z_{acc}$ that satisfies the equation $\dot H + H^2=0$ and the actual Hubble factor $H_0$ allow us to express the coefficients $b_i$ in equations (\ref{10}) as $b_1 = 3H_0^2 - b_2$ and $b_2 = 3H_0^2(2 - 3\gamma_s)/ [2 - 3\gamma_s + (1+ z_{acc})^{3(1-\gamma_s)}]$ so that the Hubble function, reads

\begin{eqnarray}
\n{14}
H(z)= \frac{H_0(1+ z_{acc})^{3/2}}{\sqrt{2-3\gamma_s+(1+ z_{acc})^{3(1-\gamma_s)}}} \times  \ \ \ \ \ \ \ \ \ \ \ \  \\
\nonumber
 \ \ \ \ \ \ \ \ \ \ \ \   \sqrt{\frac{(1+z)^{3\gamma_s}}{(1+z_{acc})^{3\gamma_s}}+(2-3\gamma_s)\frac{(1+z)^3}{(1+z_{acc})^3}}
\end{eqnarray}

\begin{figure*}
  \resizebox{20pc}{!}{\includegraphics{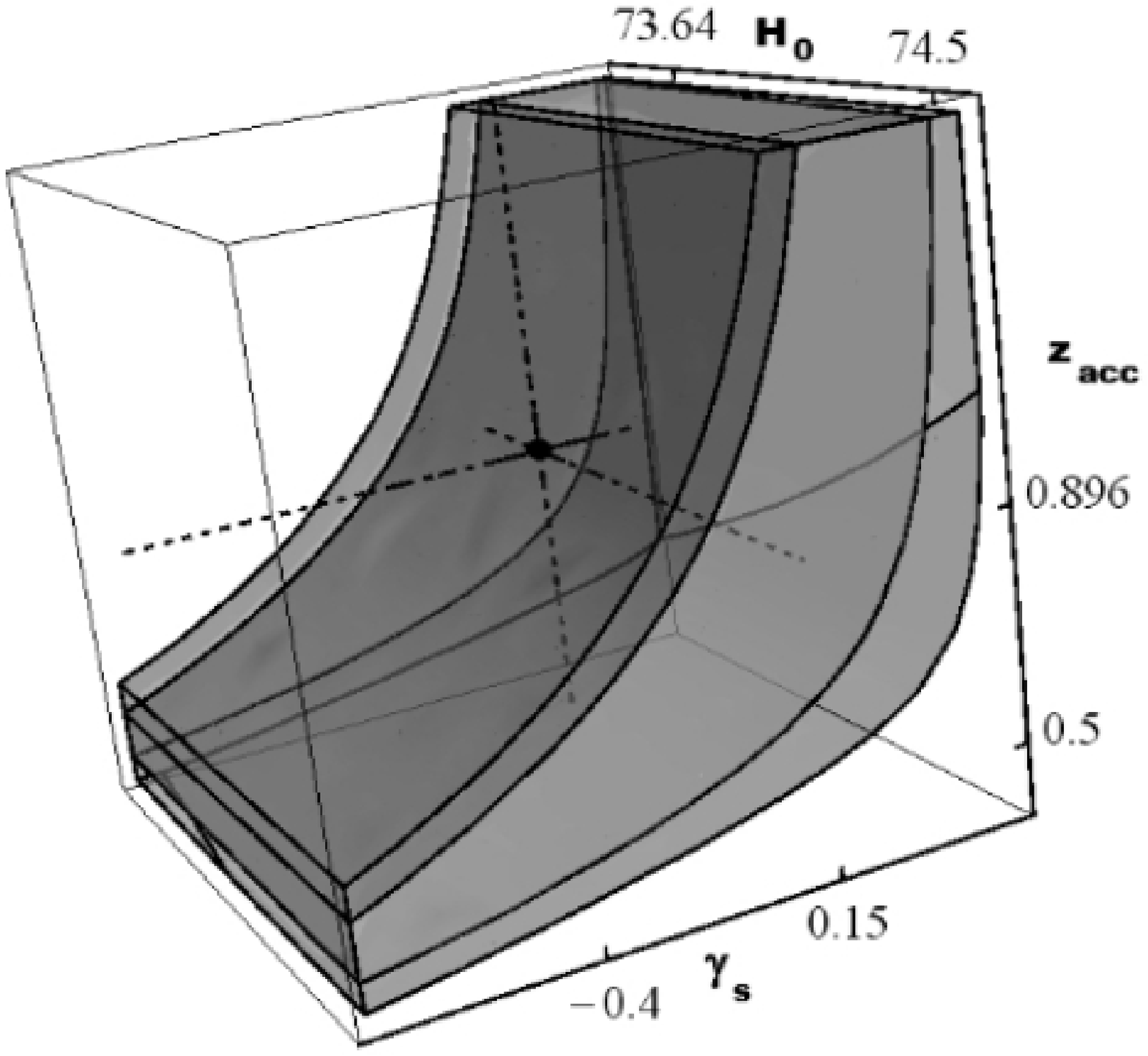}}
  \resizebox{18pc}{!}{\includegraphics{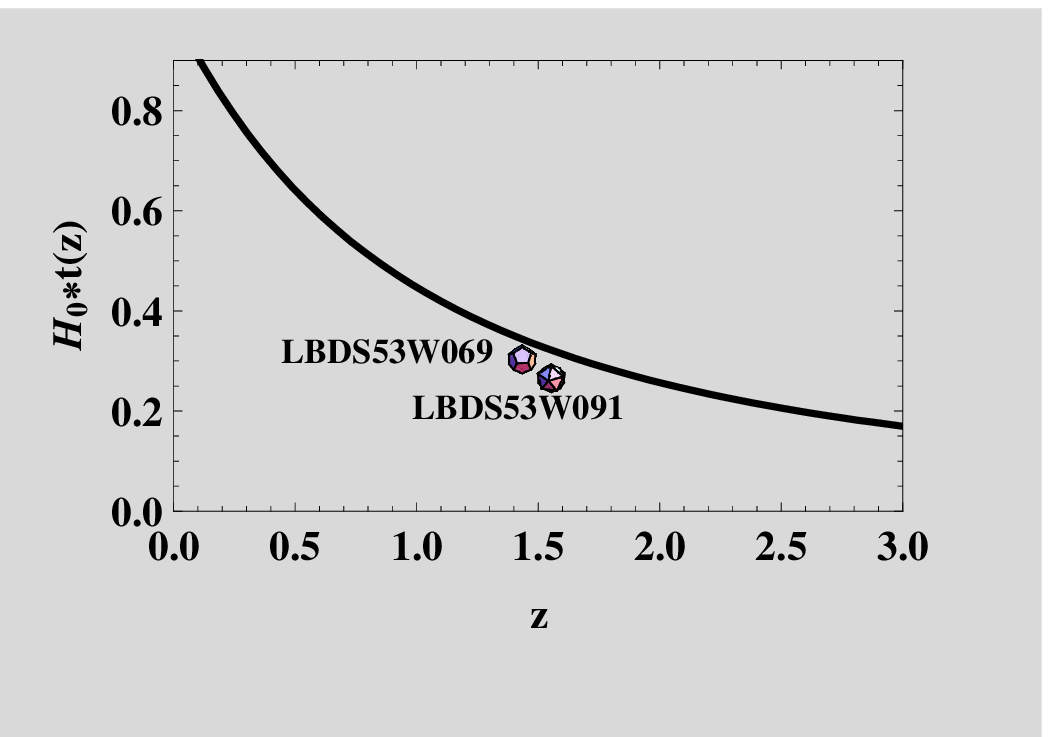}}
\caption{Left panel: Three-dimensional regions of confidence $1\sigma$ and $2\sigma$ for  $H_0$, $\gamma_s=1+\om_s$  \  and  $z_{acc}$ parameters. Right panel: Parametric curve of cosmological time $t(z)$ drawn in units of $H_0^{-1}$ for the best values $z_{acc}=0.89$, $H_0=73.6$km/sMpc  \ and  $\gamma_s=0.158$ from (\ref{tiempo}).}
\label{figura}
\end{figure*} 

We apply the $\chi^{2}$--statistical method to the Hubble data \cite{Stern:2009ep} for  constraining the cosmological parameters of the Hubble function (\ref{14}). The three-dimensional confidence regions $1\sigma$ and $2\sigma$ are shown in the left panel of Fig.\ref{figura} where the sphere indicates the best fit values $z_{acc}=0.89$, $H_0=73.6$km/sMpc  \ and  $\gamma_s=0.158$ with a  minimum value of the $\chi^{2}$ function per degree of freedom $\chi^2_{dof}=0.846$.

Interesting in the sense that we now know the values $z_{acc}$ and $H_0$ of the current Hubble parameter predicted by our model, nevertheless this information does not allow to determine the values of $\alpha $ and $\beta$ best fitted to the observational data. We must bear in mind that a feasible model of the dark sector has dark components with positive definite energy densities, accelerated expansion and non phantom dark energy. These requirements are fulfilled when    $b_1$ and $b_2$  are positive constants, which correspond to  $\alpha \geq 1$  and $ 0 \leq \beta < 2/3$.  To determine the most acceptable ranges of the parameters $\alpha$ and $ \beta$ we note that this constants are involved in the expressions of the partial energy densities $\rho_c$ and $\rho_x $ and so, we apply  the $\chi^{2}$--statistical method as above but now using the expressions

\begin{eqnarray}
\n{15}
H^2(z)=  H_0^2\frac{(\om_s - \om_0)(1+z)^3 + \om_0(1+ z)^{3\gamma_s}}{\om_s} \ \ \ \ \ \ \ \ \  \\
\nonumber
\om_0 = \alpha\Omega_{c0} + \beta\Omega_{x0} - 1, \quad  \Omega_{i0} = \frac{\rho_{i0}}{3H_0^2}, \quad \Omega_{c0} + \Omega_{x0} = 1. 
\end{eqnarray}

\begin{table}[htbp]
\begin{center}
\begin{tabular}{|r|c|c|c|}
\hline
$[\al, \beta]$&$\Omega_{c0}$&$\om_{s}$&$\chi^2_{dof}$\\
\hline\hline
$[1.15, 0.01]$&0.30&-0.84&0.761\\
\hline
$[1.3, 0.01]$&0.27&-0.84&0.761\\
$[1.3, 0.1]$&0.21&-0.84&0.761\\
$[1.3,0.2]$&0.14&-0.84&0.761\\
$[1.3, 0.3]$&0.05&-0.84&0.761\\
\hline
$[4/3, 0.01]$&0.26&-0.84&0.761\\
$[4/3, 0.1]$&0.21&-0.84&0.761\\
$[4/3,0.2]$&0.14&-0.84&0.761\\
$[4/3, 0.3]$&0.06&-0.84&0.761\\
$[4/3, 1]$&$2.3\times10^{-9}$&-0.80&22.23\\\hline
$[1.4, 0.01]$&0.25&-0.84&0.761\\
$[1.4,0.1]$&0.19&-0.84&0.761\\
$[1.4,0.2]$&0.13&-0.84&0.761\\
$[1.4, 0.3]$&0.05&-0.84&0.761\\
\hline
\end{tabular}
\end{center}
\caption{\scriptsize{Best fit values of $\Omega_{c0}$ and $\om_s$ for a given pair of $(\al, \beta)$.}}
\label{tab:1}
\end{table}

The results of this procedure, included in Table \ref{tab:1}, show that  the holographic case $\alpha =4/3$ and $ \beta = 1$ has a  very poor statistical adjustment $\chi^2_{dof} = 22.23$,  whereas  inversely, the models with $\alpha = 4/3$ and $ \beta < 0.1$, that is $0.25R < \rho_x < 0.27R$, behave reasonably well leading to  $\chi^2_{dof} = 0.761 < 1$. 
The constants $b_1$ and $b_2$ have two sets of expressions, as they are written in terms of $z_{acc}$ and $H_0$, used in (\ref{14}), or in terms of the current parameters of density $\Omega_{c0}$ and $\Omega_{x0}$, as used in (\ref{15}). 
These sets allow to express the constant $\om_0$  as 
\be
\n{16}
\om_0 =\frac{\om_s} {1-(1+3\om_s)(1+z_{acc})^{3\om_s}},
\ee   
\no and verify that the first line of the Table \ref{tab:1} gives the correct values of $\alpha$  and $\beta$. In the next subsections, we will used these values  $\alpha = 1.15$, $\beta = 0.01$  and $\Omega_{c0} = 0.3$ in the figures and expressions for the partial densities and their ratio.


\subsection{The crisis of the age}

The age of the universe in units of $H_0^{-1}$ can be obtained as a function of the redshift $z$ with the expression 
\be
\n{tiempo}
t(z)= \int_z^{\infty} \frac{d\nu}{(1+\nu)H(\nu)}
\ee

 We depict this age-redshift relation in the right panel of Fig.\ref{figura}. The parametric curve of cosmological time $t(z)$ is drawn from (\ref{tiempo}) in units of $H_0^{-1}$ for the best values $z_{acc}=0.89$, $H_0=73.6$km/sMpc  \ and  $\gamma_s=0.158$.  Because the cosmological constraints with the Hubble data only cover redshifts over the range $0 \leq z < 2$, the comparison with cosmic milestones will be trustworthy in this range only, and for that reason we consider only  two old stellar sources such as the $4 {~\rm Gyr}$ old galaxy LBDS 53W069 at redshift $z = 1.43$ \cite{Dunlop:1997} and  the $3.5 {~\rm Gyr}$ old galaxy LBDS 53W091 at redshift $z = 1.55$ \cite{Dunlop:1996mp}. We find that at low redshift $ z < 2$, the  Ricci-like holographic dark energy model seems to be free from the cosmic-age problem, namely,   the universe cannot be younger than its constituents.

\vskip0.2cm 
\subsection{The magnitude-redshift relation}

It is well known that observations of type Ia supernova(SNe Ia) have predicted  and   confirmed that our universe is passing through an accelerated phase of expansion. Since then, the observational data  coming from these standard candles have been taken  seriously.  It is commonly believed that measuring both, their redshifts and apparent peak flux, gives a direct measurement of their luminosity distances and thus SNe Ia data provides the strongest constraint on the cosmological parameters. The theoretical distance modulus is defined as 
\be
\n{mut}
\mu(z) =5~\log_{10} {\cal D}_{L}+\mu_{0}
\ee
where $\mu_{0}=43.028$, and ${\cal D}_{L}$ is the Hubble-free luminosity distance, which for a spatially flat universe can be recast as 

\be
\n{mut'}
{\cal D}_{L}(z) =(1+z)H_{0}\int^{z}_{0}\frac{dz'}{H(z')}
\ee

\begin{figure*}
  \resizebox{20pc}{!}{\includegraphics{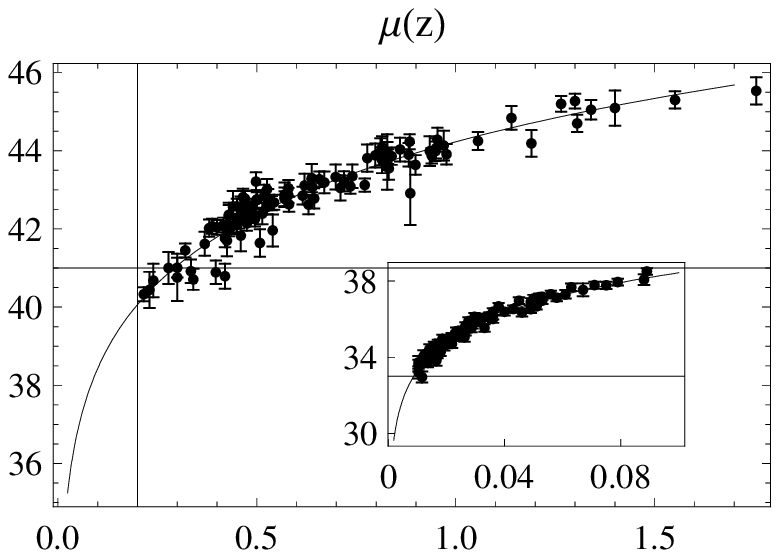}}
  \resizebox{18pc}{!}{\includegraphics{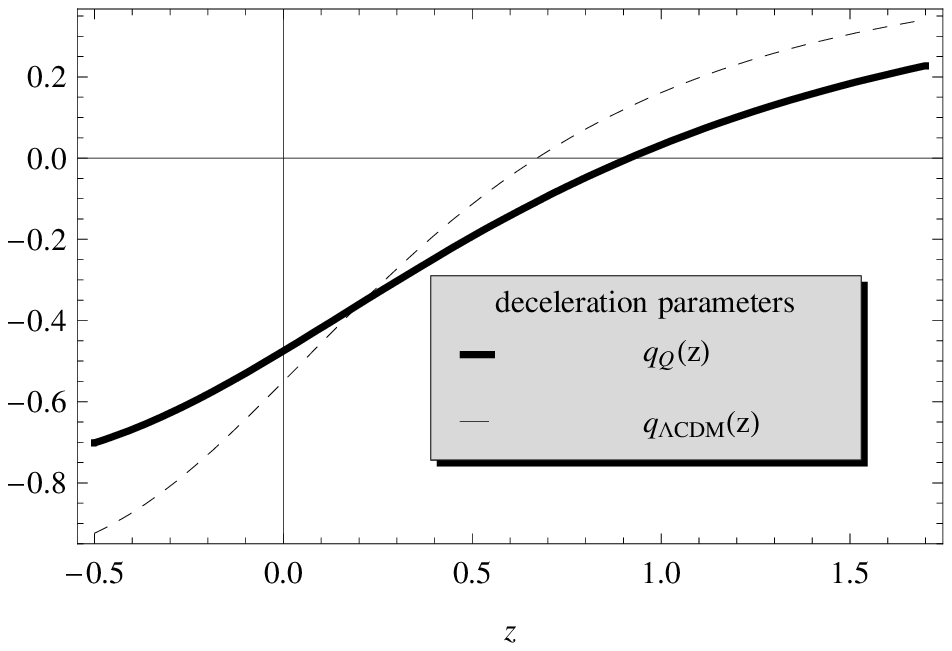}}
\caption{\scriptsize{Left panel: The predicted distance modulus (solid line) is plotted against the redshift for the interacting model. The observational data (circled point) was taken from Riess \cite{Riess:2009pu}. Right panel: The deceleration parameters $ q(z)$ (solid line) and  $q_{\Lambda CDM}(z)$ (dashed line) are compared for models with $ \Omega_c =0.3$, $\om_0=-0.65$ and $\om_s=-0.84$.}}
\label{ModuloYqes}
\end{figure*}

Using the best fit  values of $\om_s$ and $z_{acc}$ in Eqs.(\ref{14})-(\ref{mut'}) we get the theoretical distance modulus  $\mu(z)$ that we draw in the left panel of Fig.\ref{ModuloYqes} together with the observational data $\mu_{obs}(z_i)$\cite{Riess:2009pu} and their error bars. The theoretical distance modulus (\ref{mut}) will  strongly depend on the model used so taking into account a particular cosmology and  comparing its $\mu(z)$ with $\mu_{obs}(z_i)$ one can judge the plausibility of the cosmological model. As we see from Fig.\ref{ModuloYqes} our model shares an  excellent agreement with the observational data in the zones corresponding to small redshift [$ z \leq 0.1 $] and large  redshift [$0.1 \leq z \leq 1.5$]. 

\begin{figure*}
  \resizebox{20pc}{!}{\includegraphics{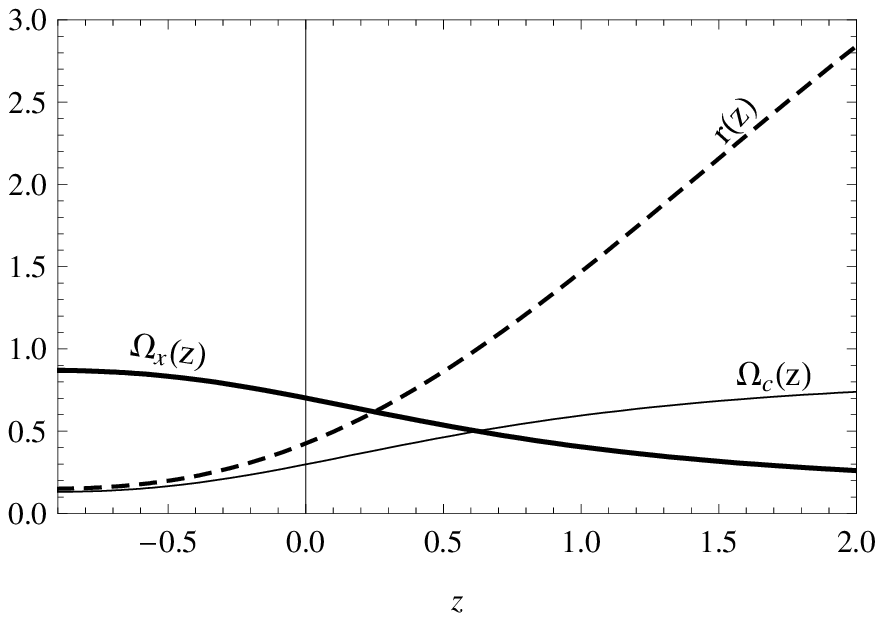}}
  \resizebox{18pc}{!}{\includegraphics{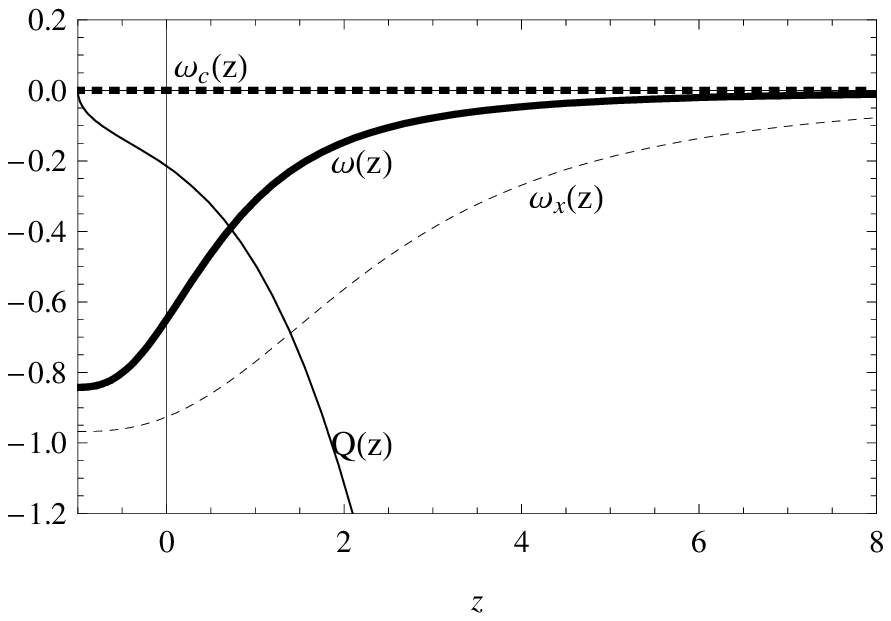}}
\caption{\scriptsize{Left panel: Density parameters $ \Omega_c$ and $\Omega_x $ and ratio $r(z)$.  Right panel: Equations of state for dark matter $\om_c$, dark energy $\om_x$ and global fluid $\om$. All the curves are drawn for the best values $ \alpha =1.15$, $ \beta =0.01$, $\om_0=-0.65$ and $\om_s=-0.84$.}}
\label{densYEos}
\end{figure*}

\vskip0.2cm 
\subsection{The deceleration parameter, equations of states and density parameters}

There are magnitudes that do not depend explicitly on the pair of constants $(\alpha,\beta)$  which selects one particular form for the energy density of the dark energy $ \rho_x =(2\dot H + 3\alpha H^2)/(\alpha-\beta)$, but on the linear combination $\om_0$ defined in (\ref{15}).  These are: the total energy $\rho$, the deceleration parameter  $q$ and the global  equation of state $\om$, whose explicit expressions can be written in terms of the transition redshift $z_{acc}$ and the asymptotic equation of state $\om_s$,  by (\ref{14}) and the functions

\be
\n{decelerat}
q(z)=-1+\frac{3}{2} \frac{(\om_s - \om_0)+\om_0(1+\om_s)(1+z)^{3\om_s}}{(\om_s - \om_0)+\om_0(1+z)^{3\om_s}},
\ee
\be
\n{w}
\om(z) = \frac{2q(z) - 1}{3}
\ee
\no with  $\om_0$ given by (\ref{16}).  
In the right panel of Fig.\ref{ModuloYqes} the deceleration parameters for all our like-holographic models with  $\om_0=-0.65$ and $\om_s=-0.84$, $ q(z)$ (solid line),  are compared with the deceleration parameter of the $\Lambda CDM$ model $q_{\Lambda CDM}(z)$ (dashed line) that holds $\Omega_c =0.3$. There we can see that the deceleration parameter of our models vanishes near $z_{acc}=0.84$, so these universes enter in the accelerated phase  more earlier than the $\Lambda$CDM model with actual density parameters $\Omega_{c0}=0.3$ and $\Omega_{x0}=0.7$. 
The effective equation of state $\om$, is plotted in the right panel of Fig.\ref{densYEos} and looking there, we conclude that our models have $ -1< \omega(z) < 0$ in the interval $z\geq 0$. More precisely,  $\omega(z)$ begins like non-relativistic matter, decreases rapidly around $z=2$ and ends with the asymptotic value $\omega_{s}=-0.84$.

Instead, the density parameters $\Omega_c =\rho_c/3H^2$ and $\Omega_x=\rho_x/3H^2$, their ratio $r =\Omega_c/\Omega_x$, the equation of state for the dark energy $\om_x$ of Eq.(\ref{4}) and also the interaction used $Q$ of Eq.(\ref{8}), are described explicitly in terms of $\alpha$ and $\beta$ by the expressions

\be
\n{Omegacdez}
\Omega_c(z)=\frac{(1- \beta)(\om_s - \om_0)+\om_0(\om_s + 1 - \beta)(1+z)^{3\om_s}}{\Delta [(\om_s - \om_0)+\om_0(1+z)^{3\om_s}]}
\ee

\be
\n{OmegaXdez}
\Omega_x(z)=\frac{(\alpha - 1)(\om_s - \om_0)+\om_0(\alpha - 1 - \om_s)(1+z)^{3\om_s}}{\Delta[(\om_s - \om_0)+\om_0(1+z)^{3\om_s}]}
\ee
\be
\n{rdez}
r=\frac{\om_{0}(\beta-\om_s-1)(1+z)^{3\om_s}+(1-\beta)(\om_0-\om_{s})}{\om_{0}(\om_s+1-\al)(1+z)^{3\om_s}+(\al-1)(\om_{0}-\om_{s})},
\ee
\be
\n{wxdez}
\om_x(z) = (\alpha - 1)r(z)+ \beta - 1,
\ee
\be
\n{26}
Q=\frac{(\alpha\beta-1-\om_s)\rho+(\alpha+\beta-2-\om_s)\rho'}{\alpha - \beta}
\ee

The density parameters $ \Omega_c$ and $\Omega_x $ and their ratio $r(z)$ are plotted in the left panel of Fig.\ref{densYEos} for the best values $ \alpha =1.15$, $ \beta =0.01$, $\om_0=-0.65$ and $\om_s=-0.84$, where we can see that the general linear interaction $Q$ helps to alleviate the coincidence problem. The later is drawn in the right panel of Fig.\ref{densYEos} together with the dark energy equation of state $\omega_x$ which has $ -1< \omega_x(z) < 0$ for $z\geq 0$.
The linear interaction $Q$ of Eq.(\ref{26}) corresponds to the choice $u =  \alpha \beta -1-\om_s$ in the Eq.(\ref{8}) and its curve  is always negative satisfying the second law of thermodynamics that requires the energy flow goes from dark energy to dark matter \cite{Pavon:2007gt}.


\section{Conclusions}
 We have examined a modified holographic Ricci dark energy  coupled  with cold dark matter and found that this scenario describes satisfactorily the behavior of the energy densities of both dark components alleviating the problem of the cosmic coincidence. 
We have shown that the compatibility between the modified and the global conservation equations  restricts  the equation of state of the dark  energy component  relating it to the ratio of energy densities. This constrain makes the holographic density always interacts with the non-holographic component  except in the unlikely event that $\alpha = 1$, which  is forbidden for positive energy densities.  From the observational point of view we have  obtained the best fit values of the cosmological parameters $z_{acc}=0.89$, $H_0=73.6$km/sMpc  \ and  $\gamma_s=0.158$   with   a $\chi^{2}_{dof}=0.761 < 1$ per degree of freedom. The $H_{0}$ value is in agreement with the reported in the literature \cite{Riess:2009pu} and the critical redshift $z_{acc}=0.89$ is consistent with BAO and CMB data \cite{Li:2010da}. We have found that in the redshift interval where is trustworthy compared with old stellar sources the model is free from the cosmic-age problem.

\begin{acknowledgments}

The authors are grateful to the I CosmoSul' organizers for their kindness and good organization. LPC thanks  the University of Buenos Aires under Project No. X044 and the Consejo Nacional de Investigaciones Cient\'{\i}ficas y T\' ecnicas (CONICET) under Project PIP 114-200801-00328 for the partial support of this work during their different stages. MGR is partially supported by CONICET.

\end{acknowledgments}

\end{document}